%% file: manuscript.tex
\documentclass[10pt,aps,prl,amsmath,amssymb,twocolumn,superscriptaddress,floatfix]{revtex4-2}

\usepackage{graphicx}
\usepackage{braket}
\usepackage{bm}
\usepackage[bookmarksnumbered,pdfpagelabels=true,plainpages=false,colorlinks=true,linkcolor=blue,citecolor=blue,urlcolor=blue]{hyperref}
\usepackage{bbm}
\usepackage{cleveref}
\usepackage{xcolor}
\usepackage[normalem]{ulem}
\input{defs.tex}

\crefname{figure}{Fig.}{Figs.}
\Crefname{figure}{Fig.}{Figs.}

\crefname{equation}{Eq.}{Eqs.}
\Crefname{equation}{Eq.}{Eqs.}

\begin{document}

\title{Regularized Neural Quantum States for Heisenberg Magnets}

\def\afflux{Department of Physics and Materials Science, University of Luxembourg, L-1511 Luxembourg, Luxembourg}

\author{\v{S}tefan Li\v{s}\v{c}\'{a}k}
\email{stefan.liscak@uni.lu}
\affiliation{\afflux}
\author{Vladyslav M. Kuchkin}
\affiliation{\afflux}
\author{Andreas Michels}
\affiliation{\afflux}
\author{Thomas L. Schmidt}
\email{thomas.schmidt@uni.lu}
\affiliation{\afflux}
\author{Andreas Haller}
\email{andreas.haller@pe-systems.de}
\affiliation{\afflux}
\affiliation{PE-Systems GmbH, Landwehrstraße 55, 64293 Darmstadt, Germany}

\date{\today}

\begin{abstract}
We introduce an approach to solving locally ordered quantum spin systems using neural network quantum states.
First, our approach rests on choosing a spin basis constructed from the magnetic texture of a classical or quantum reference state.
Second, the method improves convergence and suppresses parameter-space drift through a regularized Riemannian gradient descent adapted to the quantum state geometry of the variational state manifold.
Combining these techniques enables stable ground-state calculations for two- and three-dimensional chiral spin systems with Dzyaloshinskii-Moriya interactions.
The resulting variational ground-state energies are competitive with those of high-fidelity density matrix renormalization group calculations and, for periodic and layered geometries, provide significantly tighter upper bounds on the ground-state energy.
\end{abstract}
\maketitle

\begin{figure}[t]
    \centering
    \includegraphics[width=\columnwidth]{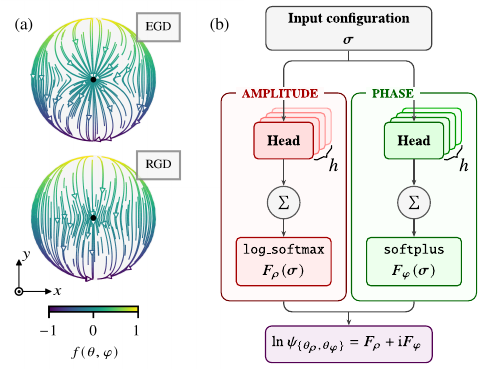}
    \caption{
        Geometric interpretation of gradient descent and the NQS architecture.
        (a)~Schematic illustration of gradient flows on the sphere $\mathbb{S}^2$ (top view), directed toward the minima of the objective function $f(\theta, \varphi)=\sin^2{\theta}\sin{\varphi}$.
        Euclidean gradient descent (EGD) defines steepest descent with respect to the identity metric in the angular coordinates ($g=\mathbbm 1$), whereas Riemannian gradient descent (RGD) uses the induced spherical metric ($g={\rm diag}(1,\sin^2\theta)$).
        Consequently, EGD is not intrinsically adapted to the spherical geometry and may exhibit coordinate-dependent behavior near the poles (black dot), whereas RGD accounts for the curvature of the sphere when determining the update direction and size.
        (b)~Wavefunction ansatz architecture.
        The input configuration $\bsigma$ is processed by $h$ parallel heads in separate amplitude and phase branches.
        Their outputs are aggregated to obtain the real-valued
        log-amplitude $F_\rho(\bsigma)$ and phase $F_\varphi(\bsigma)$, which define the output $\ln\psi_{\btheta}(\bsigma)=F_\rho(\bsigma)+\ri F_\varphi(\bsigma)$.
    }
    \label{fig:method_overview}
\end{figure}

\textit{Introduction}.--
Finding ground states of interacting quantum systems remains a challenging task because of the exponential complexity of generic many-body wavefunctions.
In recent years, Neural Quantum States (NQS) have emerged as a powerful variational ansatz, leveraging the expressivity of artificial neural networks to represent complex wavefunctions \cite{Troyer2017,Deng2017,Choo2018,Lange2024,Medvidovic2024}.
While standard network optimization methods treat variational parameters as coordinates on a flat space \cite{Kingma2017}, quantum variational states naturally reside on a curved manifold, where distances are dictated by the quantum metric \cite{Provost1980,Amari1998}.
This observation underlies Riemannian-gradient methods, such as Stochastic Reconfiguration (SR) \cite{Sorella1998,Troyer2017,Pervez2025}, which improve the optimization by following the geometry of the variational manifold [see \cref{fig:method_overview}(a)].
However, small eigenvalues of the quantum metric, arising from redundant or weakly resolved parameter directions and finite-sampling noise, can lead to unstable optimization, particularly for large and deep networks \cite{Rende2024,Chen2024,GuzmanCordero2025,Liu2026IRL}.
Diagonal regularization of the metric is therefore commonly introduced.
Beyond improving numerical conditioning, the diagonal shift acts as a statistical filter on finite-sample SR updates \cite{Hur2026}.
Restricting optimization to low-dimensional parameter subspaces has also been shown to improve NQS training stability \cite{Wei2026}.
Nevertheless, regularizing the metric does not directly constrain parameter drift along redundant directions.
In particular, parameters can drift along nearly flat directions of the energy landscape without producing a corresponding change in the physical quantum state.
They can thereby grow until they exceed numerical precision \cite{Solinas2026}.

In this work, we introduce a Regularized Riemannian Gradient Descent (RRGD) scheme that controls this parameter-space drift while retaining the quantum geometry of the variational manifold.
We derive the update from a Lagrangian with simultaneous constraints on the quantum distance of the update and the parameter norm at its endpoint.
The resulting soft constraint suppresses drift along redundant parameter directions while preserving the Riemannian structure of the variational update.
Furthermore, basis rotations provide a complementary route to improving NQS optimization.
Previous studies have demonstrated that global basis choices can modify the variational representation and the parameter-space distance between the initial and target states~\cite{Pei2021, Kozic2026, CortesSantamaria2026,Paul2026}.
Additionally, local orbital rotations have been jointly optimized with NQS parameters to improve the variational accuracy and the optimization landscape~\cite{Moreno2023}.
Here, we do not treat the basis transformation as an additional set of variational parameters.
Instead, we use a physically informed, fixed local basis for ordered Heisenberg spin-$1/2$ magnets, which avoids introducing further redundant variational degrees of freedom.
This provides a complementary approach to the optimization problem: the local frames improve the starting point, and RRGD regularizes the subsequent evolution through parameter space.
We demonstrate that their combination enables stable and fast optimization of shallow NQS architectures in single precision for various atomistic spin-$1/2$ chiral models known to host skyrmion textures~\cite{Haller2022,Joshi2023,Corte2026,Liscak2026}.
Our simulations are implemented in a custom JAX-based code, leveraging composable automatic differentiation and just-in-time compilation \cite{jax2018github}.
We refer to the Supplemental Material \cite{Suppl} and~\cite{Novak2022} for additional details on the network and the training method.

\textit{Network Layout and Initialization}.--
Throughout this work, we represent the many-body wavefunction using an NQS, which assigns a complex amplitude $\psi_{\bm\theta}(\bsigma) = \braket{\bsigma | \psi_\btheta}$ to each configuration $\bsigma = (\sigma_1, \dots, \sigma_N)$ in a chosen computational basis $\{ \ket{\bsigma} \}$ \cite{Troyer2017,Sharir2022}.
We parameterize its logarithmic amplitude $\rho$ and phase $\varphi$ using separate shallow feed-forward neural networks (FFNNs),
\begin{align}
    \ln\psi_{\bm \theta}(\bsigma)
    =
    \rho_{\bm\theta_\rho}(\bsigma)
    +
    \ri\varphi_{\bm\theta_\varphi} (\bsigma),
    \label{eq:wf_decomp}
\end{align}
where $\btheta = \{ \bm\theta_\rho,\bm\theta_\varphi \} \in \mathbb{R}^{n_{\rm p}}$ represents the variational parameters.
Both subsets are optimized simultaneously, as in recent phase-resolved NQS studies of frustrated and chiral magnets \cite{Westerhout2019,Roth2023,Joshi2024}.
For the chosen parameterization, the quantum metric is block diagonal in $(\btheta_\rho,\btheta_\varphi)$.
Each network consists of $h$ parallel input transformations (heads)~\cite{Suppl}.
This architecture increases expressivity without deeply stacked network structures [see \cref{fig:method_overview}(b)], thereby sidestepping the optimization difficulties associated with deep architectures in the sign-sensitive regime \cite{Chen2024,Nutakki2025,GuzmanCordero2025}.
Because variational parameters generally lack a direct physical interpretation, constructing a physically informed initialization requires understanding the state represented at a particular point in the parameter space~\cite{Deng2017,Medvidovic2024}.
For the architecture used here, the network output is configuration-independent at vanishing parameters, $\psi_{\bm \theta = \bm 0}(\bsigma)=\mathcal{Z} \neq 0$.
Consequently, in the canonical $z$-basis, the zero-parameter state is the equal-amplitude superposition
\begin{align}
    \ket{\psi_{\bm\theta=\bm0}}
    =
    {\mathcal Z}\sum_{\bm\sigma}\ket{\bm\sigma}
        =
        {\mathcal Z}\bigotimes_{i=1}^N
    \sum_{m_i=-s}^s \ket{s, m_i}_z,
\end{align}
which, for spin-$1/2$ systems, coincides with the $\hat{{\bm x}}$-polarized spin-coherent state up to normalization.
Thus, the origin of the parameter space has a direct physical interpretation, and displacements from $\btheta=\bm0$ explore the surrounding variational manifold.
However, these displacements need not correspond one-to-one to physically distinct states.
Redundant parameterizations and gauge degrees of freedom can give rise to directions along which the physical state remains unchanged, a feature shared with overparameterized NQS and transformer wavefunctions \cite{Rende2025_1,Rende2025_2,Rende2025_3}.
These redundancies are treated and partially alleviated by the RRGD algorithm.

\textit{Canonical Local Frames}.--
For the systems considered here, we use the atomistic spin-$1/2$ Hamiltonian
\begin{equation}
    \hat H
    =
    \sum_{\langle i,j\rangle} \sum_{\alpha,\beta}
    \mathcal{J}_{ij}^{\alpha\beta} \hat{S}_i^\alpha \hat{S}_j^\beta
    +
    \sum_i B \hat{S}_i^z,
\end{equation}
where $\hat{\bm{S}}_i = \hat\bsigma_i/2$ denotes the spin operator at site $i$ and $\hat\bsigma$ the vector of Pauli matrices.
The first sum runs over unique nearest-neighbor pairs.
The coupling matrix
$\mathcal{J}_{ij}^{\alpha\beta}
    =
    J \delta^{\alpha\beta}
    +
    \varepsilon^{\alpha\beta\gamma} D_{ij}^\gamma$
contains the isotropic Heisenberg exchange $J$ and the Dzyaloshinskii--Moriya interaction (DMI) $\bm{D}_{ij}$, while $B$ is a uniform external magnetic field.
Depending on its orientation relative to the normalized bond vector $\hat{\bm{r}}_{ij}$, the DMI vector takes one of two forms: $\bm{D}_{ij}=D\hat{\bm{r}}_{ij}$ for Bloch-type DMI or $\bm{D}_{ij}=D\hat{\bm{z}}\times\hat{\bm{r}}_{ij}$ for N\'eel-type DMI.
We denote these two types by $D_{\rm B}$ and $D_{\rm N}$, respectively.
To provide a physically informed starting point, we rewrite the Hamiltonian $\hat H$ in a site-dependent frame aligned with a reference spin texture.
The reference texture is obtained either from a classical spin configuration or from an intermediate Matrix Product State (MPS) variationally optimized by the Density Matrix Renormalization Group (DMRG) \cite{White1992,Schollwock2005,Schollwock2011}.
The corresponding transformation $\hat U = \prod_i \hat U_i(\Phi_i, \Theta_i)$ defines the unitarily equivalent Hamiltonian $\hat H' = \hat U^\dagger \hat H \hat U$, on which the NQS is optimized \cite{CortesSantamaria2026,Kozic2026}.
The local angles $(\Theta_i, \Phi_i)$ are determined by the reference spin at site $i$.
We choose the rotation convention such that the reference spins are mapped onto the $+\hat{\bm x}$ axis; consequently, the zero-parameter NQS represents the reference texture \cite{Suppl}.
Furthermore, the local rotation matrices are kept fixed as nontrainable hyperparameters.
Since their orientations are already determined by the reference configuration, optimizing the frame angles would enlarge the parameter space and may introduce redundant or poorly conditioned directions, without necessarily providing additional physically motivated variational freedom \cite{Rende2025_2,Nutakki2025}.
In practice, we initialize the weights with Gaussian noise of standard deviation $10^{-7}$.

\textit{Regularized Riemannian Gradient Descent}.--
For normalized variational quantum states, the local state-space distance is determined by the quantum metric \cite{Provost1980}
\begin{align}
    g_{ij}
        =
        {\rm Re}\Braket{
        \frac{\partial\psi_{\bm\theta}}{\partial\theta_i}
        |
        \left(\mathbbm 1 - \ket{\psi_{\bm\theta}}\bra{\psi_{\bm\theta}}\right)
        |
        \frac{\partial\psi_{\bm\theta}}{\partial\theta_j}
    }.
\end{align}
Minimizing the locally linearized energy at fixed quantum distance $ds^2 = \delta\btheta^T g\delta\btheta$ gives
\begin{equation}
    g\delta\bm\theta
    =
    -\eta\partial_{\bm\theta}E(\bm\theta),
    \label{eq:riemannian_gradient_descent}
\end{equation}
which corresponds to Riemannian gradient descent \cite{Amari1998}.
In practice, the quantities entering \cref{eq:riemannian_gradient_descent} are estimated using Markov-chain Monte Carlo sampling and the resulting method is known as stochastic reconfiguration \cite{Sorella1998,Troyer2017}.
The metric is in general positive semidefinite and can become ill-conditioned due to exact null directions arising from redundant parameterizations, as well as to weakly resolved directions in its stochastic estimate.
The conventional replacement $g \rightarrow g + \epsilon\mathbbm{1}$ provides a Tikhonov-type regularization \cite{Tikhonov1963,TikhonovArsenin1977,Levenberg1944,Marquardt1963,Engl1996}.
However, it bounds only the update $\delta\btheta$ and does not prevent $\btheta$ from drifting along nearly flat directions, where the drift produces little or no change in the represented quantum state.
Such drift can drive parameter magnitudes to scales that compromise finite-precision computations, even when each individual update remains bounded \cite{Solinas2026}.
We instead constrain both the quantum distance of the update and the norm of its endpoint.
To this end, we minimize the Lagrangian
\begin{align}
    \mathcal L_{\rm RRGD}(\delta\bm\theta)
     & =
    \partial_{\bm\theta}E(\bm\theta)\cdot\delta\bm\theta
    +
    \frac\lambda2\left(\delta\btheta^T g \delta\btheta - \eta'^2\right)
    \nonumber \\
     & +
    \frac\nu2
    \left(|\bm\theta + \delta\bm\theta|^2 - \mu'^2\right)
    .
    \label{eq:L_rrgd}
\end{align}
Stationarity with respect to $\delta\bm\theta$, together with the definitions $\eta=1/\lambda$ and $\mu=\nu/\lambda$, gives
\begin{align}
    (g + \mu\mathbbm1)\delta\bm\theta
    =
    -\eta \partial_{\bm\theta}E(\bm\theta)
    -\mu\bm\theta
    ,
    \label{eq:RRGD}
\end{align}
where the shift $+\mu\mathbbm{1}$ stabilizes the metric inversion, while the accompanying term $-\mu\bm\theta$ anchors the parameters to the physically motivated zero-parameter state and suppresses the aforementioned drifts.
We refer to \cref{eq:RRGD} as the Regularized Riemannian Gradient Descent (RRGD) update.
In practice, we solve \cref{eq:RRGD} iteratively using the conjugate gradient (CG) solver provided by JAX \cite{jax2018github}, initialized with the solution from the previous iteration.
We use a relative residual tolerance of \(10^{-5}\) and impose a maximum of 128 iterations.

Although the parameter-norm penalty is isotropic, its effect is filtered by the quantum metric.
In an eigenmode of $g$ with eigenvalue $\gamma_a$, the restoring contribution is weighted by $\mu/(\gamma_a+\mu)$.
RRGD therefore mainly anchors weakly resolved directions with $\gamma_a\lesssim\mu$, while recovering SR up to corrections of order $\mu/\gamma_a$ for $\gamma_a\gg\mu$.
A finite $\mu$ nevertheless introduces a controllable regularization bias, whose spectral interpretation is discussed in the Supplemental Material \cite{Suppl}.

\begin{figure}[ht]
    \includegraphics[width=\columnwidth]{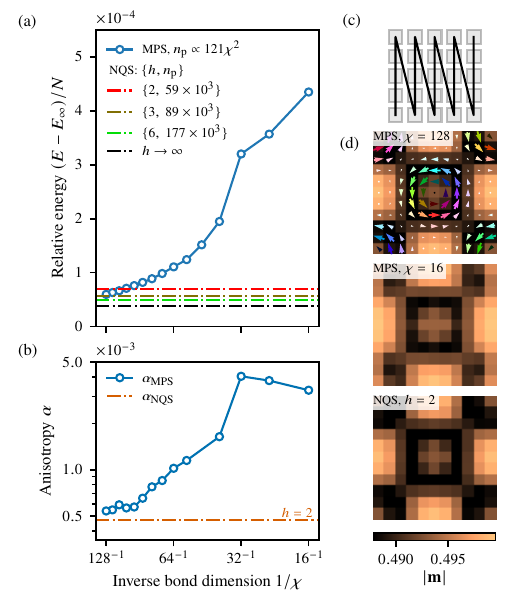}
    \caption{
        Energy convergence and symmetry properties of the skyrmion states.
        (a,b)~Bond-dimension scaling of the relative energy per site and the residual $C_4$ anisotropy of variational MPS, compared with results obtained by training $h$-headed networks (dashed lines).
        The black dashed line in panel (a) indicates the infinite-head extrapolation of the energy.
        (c)~The 1D ``zigzag'' ordering of lattice sites used for the MPS.
        (d)~Converged local spin textures for MPS with $\chi=128$, $\chi=16$ and for an NQS with $h=2$.
        Unlike the other textures, the MPS with $\chi=16$ exhibits visible vertical artifacts that break $C_4$ symmetry in the local magnetization norm $\vert \bm m \vert$.
        Throughout the paper, spin textures are color-coded using the standard scheme~\cite{Savchenko2022}: the hue is given by ${\rm atan2} (m_{y}, m_{x})$, while the saturation S and value V are determined by the $m_{z}$ component of the magnetization.
    }
    \label{fig:chiral_fm}
\end{figure}

\textit{Ferromagnetic Chiral Magnet}.--
To benchmark the NQS approach, we first consider the well-studied case of a two-dimensional skyrmion lattice \cite{Haller2022,Joshi2023} in a square-lattice chiral magnet.
We simulate an $11 \times 11$ system with periodic boundary conditions (PBC), using $J=-1$, $D_{\rm B}=-1$, and $B=-0.3$, which corresponds to a skyrmion-lattice ground state.
We use DMRG simulations to obtain an accurate estimate of the state and its quantum characteristics, and we take its energy and the distribution of the magnetic-moment length as references (see \cref{fig:chiral_fm}).
With RRGD optimization, the single-layer NQS achieves a variational energy per site of $E^{h=6}_{\mathrm{NQS}}/N=-0.659671(1)$, in proximity of but surpassing the DMRG result $E^{\chi=128}_{\mathrm{MPS}}/N=-0.6596602$.
At the same time, it approaches a lower energy variance per site compared to MPS, i.e. $\sigma^2_{\rm MPS}/N = 1.69\times10^{-4}$ vs. $\sigma^2_{\rm NQS}/N = 1.08\times10^{-4}$.
An infinite-$\chi$ extrapolation of the DMRG data yields $E_{\infty}/N=-0.6597198$.
The $h=6$ NQS energy per site lies only $4.9\times10^{-5}$ above this value, compared with $6.0\times10^{-5}$ for the $\chi=128$ MPS.
The NQS reaches this accuracy with about twenty times fewer variational parameters than the MPS ($2\chi^2 N$ vs. $2h(N+1)N$).

Although DMRG accurately describes this weakly entangled phase at sufficiently large bond dimensions, its MPS representation requires a one-dimensional mapping of the lattice [we chose the ordering sketched in \cref{fig:chiral_fm}(c)].
This 1D-chain ordering is not unique and can introduce artifacts that manifest themselves as apparent lattice symmetry breaking.
In particular, this is evident in the distribution of local spin-magnitude variations $ \vert {\bm m}_{i} \vert := \vert \braket{\hat{\bm S}_i} \vert$, which can provide a convenient proxy for the local entanglement entropy \cite{Haller2024,Liscak2026}.
Although these observables should respect the $C_{4}$ rotational symmetry of the square lattice, they generally exhibit residual anisotropy in the MPS representation for insufficiently large bond dimensions.
We quantify the resulting anisotropy through
\begin{equation}
    \alpha^2
    =
    \dfrac{1}{Ns^2}
    \sum_{i=1}^{N}
    \left(
    \vert{\bm m}_{i}\vert - \vert{\bm m}_{i^*}\vert
    \right)^2,
\end{equation}
where $i^*$ denotes the symmetry partner of site $i$ under a $\pi/2$ rotation about the center of the lattice.
As shown in \cref{fig:chiral_fm}(b,d), low-$\chi$ MPS states develop anisotropic spin-norm fluctuations aligned with the 1D path, while increasing $\chi$ progressively reduces the residual $C_4$ anisotropy.
The locally rotated NQS exhibits a much smaller $C_4$ anisotropy than MPS representations, even at large bond dimensions (see \cref{fig:chiral_fm}), which demonstrates the dimensionality-agnostic nature of the FFNN ansatz.

\begin{figure}[ht!]
    \includegraphics[width=\columnwidth]{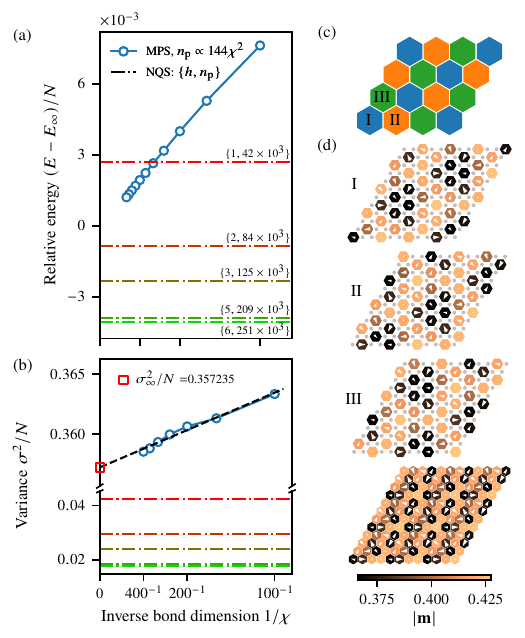}
    \caption{
        Comparison of MPS and NQS results obtained for the triangular-lattice antiferromagnetic phase.
        (a,b)~MPS bond-dimension scaling compared with NQS parallel-head scaling for the relative energy and the variance per site.
        (c)~Ordering of sublattices within the triangular lattice.
        (d)~Local magnetization exhibiting skyrmion-like structures on the individual sublattices $\mathrm{I}$, $\mathrm{II}$, and $\mathrm{III}$ and in the composite full spin texture.
    }
    \label{fig:chiral_afm}
\end{figure}

\textit{Antiferromagnetic Chiral Magnet}.--
We next consider a quantum three-sublattice antiferromagnetic chiral magnet on a $12 \times 12$ triangular lattice, which supports a skyrmion-lattice phase \cite{Corte2026}.
This phase is stabilized at $J=1$, $D_{\rm N}=0.8$, and $B=-1.76$ and, strictly speaking, requires PBC.
These, however, are challenging for conventional DMRG because of the entanglement growth in two and higher dimensions \cite{Verstraete2024}.
Although the DMRG algorithm has not reached a clear convergence plateau at the bond dimensions considered, the resulting state reproduces the three-sublattice skyrmion texture, with small finite-size distortions (see \cref{fig:chiral_afm}).
Here, however, the strongly nonconvex energy landscape makes NQS optimization sensitive to the choice of local reference frames.
A classical initialization can trap the optimization in a metastable minimum, where the NQS captures local quantum fluctuations but may fail to reach variational states of lower energy.
Such transitions may require big parameter changes, whereas Monte Carlo sampling based on local spin flips predominantly probes configurations in proximity of the current state and therefore provides little gradient information toward distant minima.
We overcome this topological trapping by a hybrid bootstrapping strategy based on a moderately converged finite-bond-dimension DMRG calculation.
The local frames are initialized from the spin orientations extracted from this intermediate-$\chi$ DMRG state, which incorporates essential nonlocal quantum correlations while remaining sufficiently close to the variational manifold explored by the NQS.
Starting from this quantum-informed reference frame, the $h$-headed NQS converges smoothly to a lower variational energy than the MPS reference simulation (see \cref{fig:chiral_afm}).

\textit{Bilayer Monoaxial Magnet}.--
To demonstrate the applicability of the NQS approach beyond strictly 2D geometries, we next consider a magnetic bilayer as an intermediate step toward systems with three-dimensional connectivity.
Specifically, we study a $10 \times 10 \times 2$ bilayer of stacked 2D monoaxial chiral magnets introduced in Ref.~\cite{Gosh2025}.
The mutually orthogonal DMI in the two layers produces a strongly frustrated magnetic texture containing localized anti-aligned (AA) points, which may be viewed as analogs of Bloch points \cite{Doring1968}.
As a result, one can expect a strong localization of quantum fluctuations near these points \cite{Kuchkin2025}.
We investigate the model at $J=-1$, $B=0$, and $D_{\rm B}=1$, with fully isotropic nearest-neighbor exchange and monoaxial intralayer DMI given by the vectors:
\begin{align}
    \bm{D}_{ij}^{\rm top} = D_\mathrm{B}\left(0, y_{i}-y_{j}\right), \quad \bm{D}_{ij}^{\rm bottom} = D_\mathrm{B}\left(x_{j}-x_{i}, 0\right),
\end{align}
where we set the lattice constant to $a=1$.

\begin{figure}[ht]
    \includegraphics[width=\columnwidth]{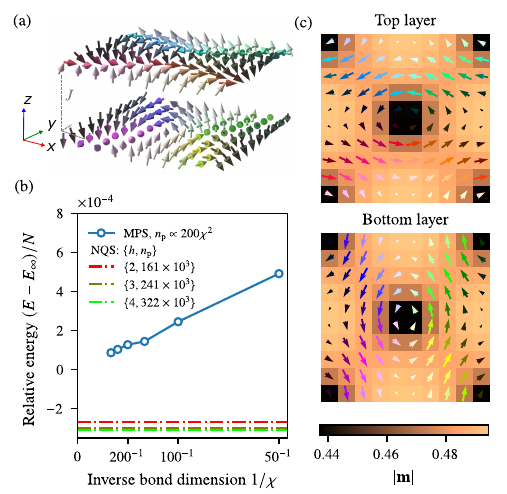}
    \caption{
        Magnetic texture and energy convergence of the coupled chiral-magnet bilayer.
        (a)~Magnetic ``checkerboard'' state stabilized in a bilayer system coupled via the Heisenberg interaction.
        The anisotropic intralayer DMI makes the spiral propagation directions in the two layers mutually orthogonal.
        (b)~Bond-dimension scaling of the energy density compared with NQS parallel-head scaling.
        (c)~Spin textures of the bilayer, showing localized AA points at the center and corners, where the spins in the two layers exhibit the strongest local anti-alignment within the obtained texture.
        This anti-alignment is reflected in the opposite saturation extremes of the arrow color-coding in the two layers.
    }
    \label{fig:chiral_bilayer}
\end{figure}

Both DMRG and NQS resolve the localized signatures of the AA points, as shown in \cref{fig:chiral_bilayer}.
The periodic in-plane geometry and the interlayer couplings make the model challenging for an MPS representation and impair DMRG convergence.
In contrast, the NQS reaches a much lower variational energy for the computational resources considered here [see \cref{fig:chiral_bilayer}(b)].
This result demonstrates that the approach is independent of the system’s interaction graph, making it directly applicable to the study of long-ranged or 3D quantum systems.

\textit{Conclusion}.--
We have presented an approach to NQS optimization that combines physically informed local frames with a regularized version of Riemannian gradient descent.
The two methods address complementary aspects of the optimization problem: the reference-aligned basis provides a physically motivated starting state at minimal parameter norm, while RRGD suppresses the parameter-space drift that accumulates along redundant directions.
This construction stabilizes the optimization even with single-precision variational parameters, thereby enabling large-scale simulations on commodity hardware.
All results reported here were obtained on a single RTX~5090 GPU in less than two GPU weeks.

We demonstrated the approach on a variety of chiral spin systems, including 2D and layered geometries.
For the square-lattice ferromagnet, we showed that a modestly-sized network can capture the 2D symmetry of local spin-magnitude fluctuations with high fidelity.
The triangular-lattice antiferromagnet highlighted the importance of reference selection: when a classical texture led to metastable optimization, local frames extracted from an intermediate DMRG calculation provided a more effective starting point.
This hybrid strategy uses only the reference spin directions and does not require the NQS to start from the exact MPS wavefunction.
In the bilayer, the NQS resolved the localized signatures of anti-aligned points, extending the demonstration beyond strictly planar geometries.
Across all benchmarks, the NQS consistently achieves variational energies competitive with or lower than those obtained by DMRG.

These results identify reference selection and parameter control as practical tools for improving NQS calculations of noncollinear quantum magnets.
Their effectiveness nevertheless depends on the reference texture and the regularization strength.
Finite anchoring can bias the optimized state, and the method does not remove the possibility of converging to metastable excitations.

The present bilayer calculations motivate extensions to a larger number of stacked layers and to fully 3D spin systems.
More broadly, encoding the classical texture in local reference frames and controlling the parameter norm through the regularized metric shift the burden from network depth toward physical insight.
This brings quantum fluctuations around large and complex spin textures within reach of shallow networks on a single GPU.

\textit{Acknowledgments}.--
The authors acknowledge financial support from the Luxembourg National Research Fund under Grants C22/MS/17415246/DeQuSky and AFR/23/17951349. V.M.K.~acknowledges financial support from the European Union’s Horizon Europe research and innovation programme under the Marie Skłodowska-Curie Grant 101203692 (QUANTHOPF).

\bibliography{references}

\clearpage
\onecolumngrid

\begin{center}
  \textbf{\large Supplemental Material:\\ Regularized Neural Quantum States for Heisenberg Magnets}
\end{center}

\setcounter{equation}{0}
\setcounter{figure}{0}
\setcounter{table}{0}
\setcounter{secnumdepth}{3}
\setcounter{section}{0}

\renewcommand{\theequation}{S\arabic{equation}}
\renewcommand{\thefigure}{S\arabic{figure}}
\renewcommand{\thetable}{S\arabic{table}}
\renewcommand{\thesection}{S\Roman{section}}
\renewcommand{\theHequation}{S\arabic{equation}}
\renewcommand{\theHfigure}{S\arabic{figure}}
\renewcommand{\theHtable}{S\arabic{table}}
\renewcommand{\theHsection}{S\arabic{section}}

\input{supplement_body}

\end{document}

%% file: defs.tex
\def\ri{\mathrm i}
\def\re{\mathrm e}

\def\bnabla{{\boldsymbol\nabla}}
\def\btheta{{\boldsymbol\theta}}
\def\bsigma{{\boldsymbol\sigma}}

\def\bO{{\boldsymbol O}}
\def\bx{{\boldsymbol x}}
\def\BC{{\mathcal A}}
\def\bBC{\boldsymbol{\mathcal A}}

\usepackage{MnSymbol}

\newcommand{\bbraket}[1]{\llangle #1 \rrangle}

%% file: supplement_body.tex
\section{Neural Network architecture}
\label{sec:nn_architecture}

The amplitude and phase networks are both implemented as shallow feed-forward architectures with a sum-of-heads structure, i.e., each one consists of $h$ parallel branches dubbed ``heads".
Specifically, for input dimension (system size) $d_{\rm in}=N$, and output width $d_{\rm out}$, each network has $h$ heads with weight matrices $W^{(i)}\in\mathbb R^{d_{\rm out}\times d_{\rm in}}$ and bias vectors ${\bm b}^{(i)}\in\mathbb R^{d_{\rm out}}$, $i=1,\dots,h$.
The first head is purely affine, with no internal activation function, and its output is therefore unconstrained:
\begin{equation}
    \bm{z}^{(1)} (\bsigma)
    =
    W^{(1)}\bm{\sigma}
    +
    \bm{b}^{(1)},
\end{equation}
while the remaining $h-1$ heads each consist of a linear map followed by a sigmoid activation $\bm{f}_\sigma(\bx)= 1 / (1 + e^{-\bx})$ applied component-wise, and a final rescaling $\mathbb{R}^{d_{\rm out}} \to \left(-1,1\right)^{d_{\rm out}}$:
\begin{equation}
    \bm{z}^{(i)} (\bsigma)
    =
    2\bm{f}_\sigma\left(
    W^{(i)}\bm{\sigma}
    +
    \bm{b}^{(i)}
    \right)
    - \bm{1},
    \quad i=2,\dots,h.
\end{equation}
The heads are summed, $\bm{\zeta}(\bsigma) = \sum_{i=1}^{h}\bm{z}^{(i)}(\bsigma)$, and a global activation function $\bm{f}_{\rm out}: \mathbb{R}^{d_{\rm out}} \to \mathbb{R}^{d_{\rm out}}$ is applied. Consequently, the final output is obtained by summing over the $d_{\rm out}$ components of the activated vector, giving:
\begin{equation}
    F(\bm{\sigma})
    =
    \sum_{k=1}^{d_{\rm out}}
    \left[
    \bm{f}_{\rm out}
    \left(
        \bm{\zeta}(\bsigma)
    \right)
    \right]_k,
\end{equation}
with $\bm{f}_{\rm out}=\texttt{log\_softmax}$ for the amplitude $\rho$, and
$\bm{f}_{\rm out}=\texttt{softplus}$ for the phase $\varphi$. This yields the outputs
\begin{align}
    F_\rho(\bsigma)
    & =
    \sum_{k=1}^{d_{\rm out}} \zeta_k(\bsigma) - d_{\rm out}\ln\sum_{k=1}^{d_{\rm out}} e^{\zeta_k(\bsigma)},
    \\
    F_\varphi(\bsigma)
    &=
    \sum_{k=1}^{d_{\rm out}} \ln\left(1 + e^{\zeta_k(\bsigma)}\right).
\end{align}
With this form the zero-parameter state is configuration independent. In other words, when all weights and biases are zero, we have $F_\rho=-d_{\rm out}\ln d_{\rm out}$ and $F_\varphi=d_{\rm out}\ln 2$.
Each head contributes $d_{\rm in}d_{\rm out}+d_{\rm out}$
parameters, giving each network a total of $h(d_{\rm in}+1)d_{\rm out}$ trainable parameters. For simplicity we choose the amplitude and phase networks to have symmetric architectures. The full complex wavefunction therefore has
\begin{equation}
    n_{\rm p}=2h(d_{\rm in}+1)d_{\rm out}
\end{equation}
trainable parameters.
We consider $d_{\rm out}=d_{\rm in}$, making the parameter count scale quadratically with the system size, $n_{\rm p}={\cal O}(N^2)$.

The purpose of retaining one purely affine head is to preserve a direct linear pathway from the input to the head outputs, which we found to improve convergence in the small-$h$ regime.
This is motivated by the fact that sigmoid activations have bounded derivatives and can attenuate gradients for large-magnitude preactivations, potentially slowing optimization when the number of heads is small.
We also tested a fully symmetric variant in which all heads include an internal sigmoid activation. For small $h$ (e.g. $h=1$), the fully activated variant converges slower within a fixed iteration budget, while for larger $h$ (e.g.\ $h=6$) it shows slightly improved convergence as opposed to the asymmetric one.
In both cases, however, the qualitative behavior of the optimizer and the physical conclusions remain unchanged. All results presented in the main text use the architecture with one affine head.

We estimate the energy trend with increasing network size using unweighted linear least-squares fits of the variational energy per site to $E(h)=E_{h\to\infty}+a/h$, where $h$ is the number of heads and $E_{h\to\infty}$ is the fitted intercept at $1/h=0$.
At fixed system size, the parameter count $n_{\rm p}$ is proportional to $h$, so this procedure is equivalent to a linear fit in $1/n_{\rm p}$.
Given the limited number of network sizes considered, this procedure does not establish an asymptotic $1/h$ scaling law.
The extrapolated values should therefore be understood as indicative estimates rather than controlled infinite-network-size limits.
Our benchmark comparisons are instead based on the converged finite-$h$ variational energies.

\subsection{Initialization}

For a general spin-$s$ degree of freedom, the equal-amplitude superposition of the canonical $z$-basis states does not, in general, correspond to a state polarized along $\hat{\bm x}$.
However, since having a easily interpretable reference state is convenient for the optimization, we may still choose to initialize the network to represent a fully polarized state along $+\hat{\bm x}$.
This can be done by introducing a configuration-dependent reference amplitude $\psi_{\rm ref}(\bm\sigma)$, such that the network represents the wavefunction
\begin{equation}
    \psi_{\bm\theta}(\bm\sigma)
    =
    \psi_{\rm ref}(\bm\sigma)
    \exp\left(\rho_{\bm\theta_\rho}(\bm\sigma) + \ri\varphi_{\bm\theta_\varphi}(\bm\sigma)\right),
\end{equation}
where $\psi_{\rm ref}(\bm\sigma)>0$ is a real, parameter-free function of the configuration $\bm\sigma$.
As it is untrainable, this amplitude modulation does not change the expressivity of the network nor the optimization landscape, but allows us to initialize the zero-parameter state to a desired reference state.

We look for the amplitude $\psi_{\rm ref}(\bm\sigma)$ by rotating the canonical $z$-basis $\ket{s,s}_z$ to obtain a local state maximally polarized along $+\hat{\bm x}$:
\begin{align}
    \ket{s,s}_x
     & =
    e^{-\ri\frac{\pi}{2}\hat S^y}
    \ket{s,s}_z
    \nonumber \\
     & =
    e^{\frac{\pi}{4}\left(\hat S^- - \hat S^+\right)}
    \ket{s,s}_z
    \nonumber \\
     & =
    e^{\tan\left(\frac{\pi}{4}\right)\hat S^-}
    e^{2 \ln\left(\cos\frac{\pi}{4}\right) \hat S^z}
    e^{-\tan\left(\frac{\pi}{4}\right)\hat S^+}
    \ket{s,s}_z
    \nonumber \\
     & =
    2^{-s} e^{\hat S^-}
    \ket{s,s}_z.
\end{align}
Expanding this state in the canonical $z$-basis gives
\begin{align}
    \ket{s,s}_x
     & =
    \sum_{m=-s}^{s}
    \braket{s,m | 2^{-s} e^{\hat S^-} | s,s}_z
    \ket{s,m}
    \nonumber
    \\
     & =
    \sum_{m=-s}^{s} 2^{-s}
    \sum_{n=0}^{2s} \frac{1}{n!}
    \braket{s,m |  (\hat S^-)^n | s,s}_z
    \ket{s,m}
    \nonumber
    \\
     & =
    \sum_{m=-s}^{s} 2^{-s}
    \sum_{n=0}^{2s} \left(\frac{(2s)!}{(2s-n)! n!}\right)^{\frac{1}{2}}
    \delta_{m,s-n}
    \ket{s,m}
    \nonumber
    \\
     & =
    \sum_{m=-s}^{s}
    2^{-s}
    \binom{2s}{s+m}^{\frac{1}{2}}
    \ket{s,m},
\end{align}
up to convention-dependent phase factors. Thus, for a product reference state, the configuration-dependent reference amplitude is
\begin{equation}
    \psi_{\rm ref}(\bsigma)
    =
    \prod_{i=1}^N
    2^{-s}
    \sqrt{
        \binom{2s}{s+m_i}
    },
\end{equation}
where $\bsigma=(m_1,\ldots,m_N)$.
For spin-$1/2$, the two binomial coefficients are identical and $\psi_{\rm ref}$ reduces to a configuration-independent constant, recovering the equal-amplitude initialization used in the present work.
For higher spins, the nonuniform amplitudes are therefore not an additional variational assumption, but follow directly from requiring the zero-parameter state to be fully polarized along $+\hat{\bm x}$.

\section{Local Frames and Variational Monte Carlo Estimators}

\subsection{Canonical local-frame transformation}

Let the reference spin direction at site $i$ be specified by the unit vector
\begin{equation}
    {\bm n}_i = (\sin\Theta_i\cos\Phi_i, \sin\Theta_i\sin\Phi_i, \cos\Theta_i),
\end{equation}
where $\Theta_i$ and $\Phi_i$ are the polar and azimuthal angles in the canonical $z$-basis.
We introduce a site-dependent unitary transformation $\hat{U}=\prod_i \hat{U}_i$, where
\begin{equation}
    \hat{U}_i
    =
    \exp{\left[-\ri\Phi_i \hat{S}_i^z\right]} \exp{\left[-\ri(\Theta_i - \pi/2) \hat{S}_i^y\right]}.
\end{equation}
The induced transformation of the spin-$1/2$ operators is represented by the active-rotation matrix $R_i= R^z(\Phi_i) R^y(\Theta_i - \pi/2)  \in {\rm SO}(3)$, with the convention
\begin{equation}
    \hat U_i^\dagger\hat S_i^\alpha\hat U_i
    =
    \sum_{\mu}
    R_i^{\alpha\mu}\hat S_i^\mu.
\end{equation}
For the rotation matrix elements, it holds that $R_i^{\alpha\beta}=
    \frac{1}{2}{\rm Tr}\left[\hat\sigma^\beta\hat U_i^\dagger\hat\sigma^\alpha\hat U_i\right]$.
Additionally, the shift by $\pi/2$ ensures that $R_i \hat{\bm x} = \bm n_i$. Consequently, a local-frame state polarized along $+\hat{\bm x}$ is mapped by $\hat U$ onto the reference spin texture, and the zero-parameter NQS thus represents the reference state.

The laboratory-frame Hamiltonian is written as
\begin{equation}
    \hat H
    =
    \sum_{\langle i,j\rangle}
    \mathcal J_{ij}^{\alpha\beta}
    \hat S_i^\alpha\hat S_j^\beta
    +
    \sum_i B_i^\alpha\hat S_i^\alpha,
    \label{eq:hamiltonian}
\end{equation}
where the coupling tensor $\mathcal{J}_{ij}^{\alpha\beta} = J \delta^{\alpha\beta}
    +
    \varepsilon^{\alpha\beta\gamma} D_{ij}^\gamma$ accounts for the Heisenberg exchange $J$, and the Dzyaloshinskii-Moriya interaction (DMI) $\bm{D}_{ij}$.
Depending on the type of DMI, the DMI vector is either parallel (Bloch-type) or perpendicular (Néel-type) to the displacement vector $\bm{r}_{ij} = \bm{r}_j - \bm{r}_i$ between the two spins.
We write $\bm{D}_{ij}=D\bm{r}_{ij}/|\bm{r}_{ij}|$ for Bloch-type DMI and $\bm{D}_{ij}=D\hat{\bm{z}}\times\bm{r}_{ij}/|\bm{r}_{ij}|$ for Néel-type DMI, where $D$ is the interaction strength and $\hat{\bm{z}}$ is the unit vector perpendicular to the plane of the spins.

The Hamiltonian in the local frame is
\begin{equation}
    \hat H'
    =
    \hat U^\dagger\hat H\hat U
    =
    \sum_{\langle i,j\rangle}
    \tilde{\mathcal J}_{ij}^{\mu\nu}
    \hat S_i^\mu\hat S_j^\nu
    +
    \sum_i
    \tilde B_i^\mu\hat S_i^\mu,
    \label{eq:rotated_hamiltonian}
\end{equation}
where
\begin{align}
    \tilde{\mathcal J}_{ij}^{\mu\nu}
     & =
    \sum_{\alpha,\beta}
    R_i^{\alpha\mu}
    \mathcal J_{ij}^{\alpha\beta}
    R_j^{\beta\nu},
    \label{eq:rotated_couplings} \\
    \tilde B_i^\mu
     & =
    \sum_\alpha B_i^\alpha R_i^{\alpha\mu}.
    \label{eq:rotated_field}
\end{align}
The site-dependent transformation generally produces
a fully anisotropic coupling tensor in the local frame. The matrix
elements derived below apply to this general tensor without
modification.

Let $\ket{\psi'}$ denote the network-represented state optimized for $\hat H'$. The corresponding state in the original laboratory frame is
\begin{equation}
    \ket{\psi}=\hat U\ket{\psi'}.
\end{equation}
An observable $\hat O$ defined in the laboratory frame must therefore be evaluated in the local frame using
\begin{equation}
    \hat O'
    =
    \hat U^\dagger\hat O\hat U,
    \qquad
    \frac{\braket{\psi|\hat O|\psi}}
    {\braket{\psi|\psi}}
    =
    \frac{\braket{\psi'|\hat O'|\psi'}}
    {\braket{\psi'|\psi'}}.
    \label{eq:rotated_observable}
\end{equation}
In particular, the laboratory-frame local magnetization and two-point correlations are recovered as
\begin{align}
    \Braket{\hat S_i^\alpha}_{\rm lab}
     & =
    \sum_\mu
    R_i^{\alpha\mu}
    \Braket{\hat S_i^\mu}_{\rm loc},
    \label{eq:rotated_magnetization} \\
    \Braket{\hat S_i^\alpha\hat S_j^\beta}_{\rm lab}
     & =
    \sum_{\mu,\nu}
    R_i^{\alpha\mu}
    R_j^{\beta\nu}
    \Braket{\hat S_i^\mu\hat S_j^\nu}_{\rm loc}.
    \label{eq:rotated_correlations}
\end{align}

\label{sec:matrix_elements}

\subsection{Variational Monte-Carlo Estimators}
Having defined the local-frame Hamiltonian and the corresponding transformation of physical observables, we next derive the variational Monte Carlo estimators used to optimize $\ket{\psi'}$. For notational simplicity, in the following we write $\hat H$ and $\ket{\psi}$ for the Hamiltonian and state represented in the basis in which the Monte Carlo sampling is performed. In the canonical local frames, these quantities are understood as $\hat H'$ and $\ket{\psi'}$, respectively.

Consider a quantum state expressed in a complete basis $\{\ket{\bsigma}\}$, i.e., $\ket\psi = \sum_{\bsigma}\ket{\bsigma}\braket{\bsigma|\psi} = \sum_{\bsigma} \psi(\bsigma) \ket{\bsigma}$ with a wavefunction $\psi(\bsigma)=\braket{\bsigma|\psi}$. The expectation value of an observable $\hat A$ in the basis of configurations $\bsigma$ is given by
\begin{align}
    \braket{\hat A}
     & =
    \frac{\braket{\psi | \hat A | \psi}}{\braket{\psi | \psi}}
    =
    \sum_{\bsigma,\bsigma'}
    \braket{\bsigma | \hat A | \bsigma'}\frac{\psi^*(\bsigma)\psi(\bsigma')}{\braket{\psi | \psi}}
    \nonumber \\
     & =
    \sum_{\bsigma}
    \frac{|\psi(\bsigma)|^2}{\braket{\psi | \psi}}
    \sum_{\bsigma'}
    \braket{\bsigma | \hat A | \bsigma'}
    \frac{\psi(\bsigma')}{\psi(\bsigma)}
    \nonumber \\
     & =
    \sum_{\bsigma}
    p(\bsigma)
    A_{\rm loc}(\bsigma),
    \label{eq:local_estimator}
\end{align}
where in the second line we insert $\psi(\bsigma)/\psi(\bsigma)$, and in the last line we define the sampling probability $p(\bsigma) = |\psi(\bsigma)|^2/\braket{\psi|\psi}$. In addition, we defined the local estimator associated with the observable,
\begin{equation}
    A_{\rm loc}(\bsigma) = \sum_{\bsigma'} \braket{\bsigma | \hat A | \bsigma'}\frac{\psi(\bsigma')}{\psi(\bsigma)}.
\end{equation}
Using configuration samples $\{\bsigma_i\}_{i=1}^{n_s}$ appropriately drawn from the probability distribution $p(\bsigma)$, we can approximate the sum in \cref{eq:local_estimator} according to
\begin{align}
    \braket{\hat A} \approx \frac1{n_s}\sum_{i=1}^{n_s}A_{\rm loc}(\bsigma_i)
    =
    \bbraket{
        A_{\rm loc}
    }
    ,
\end{align}
where we introduced the notation for the sample average of a general function
\begin{align}
    \bbraket{f}
    \coloneq
    \frac1{n_s}
    \sum_{i=1}^{n_s}f(\bsigma_i)
    .
    \label{eq:sample_average}
\end{align}

\subsection{Pair and single-spin interactions}
To evaluate the local estimators relevant for the evaluation of the energy and its gradient, we require the configuration-basis matrix elements of the single- and two-spin operators entering the Hamiltonian. We derive these expressions below for a general coupling tensor, so they apply directly to the transformed coefficients $\tilde{\mathcal J}_{ij}^{\alpha\beta}$ and $\tilde B_i^\alpha$.

The local spin operators are proportional to the Pauli matrices, i.e., $\hat{S}^\alpha = \frac12 \hat{\sigma}^\alpha$ with $\alpha = x,y,z$. Their action on the canonical spin-$1/2$ basis is given by:
\begin{align}
    \hat{S}^x_i \ket{\bsigma} & = \frac12 \ket{\bsigma^{(i)}},           \\
    \hat{S}^y_i \ket{\bsigma} & = \frac\ri2\sigma_i \ket{\bsigma^{(i)}}, \\
    \hat{S}^z_i \ket{\bsigma} & = \frac12\sigma_i\ket{\bsigma},
\end{align}
where $\sigma_i = \pm 1$ for the $i$-th spin of configuration $\bsigma$ being up/down, respectively, and $\ket{\bsigma^{(i)}}$ represents $\bsigma$ with the $i$-th spin flipped.

Using the action of the spin operators on the basis states, we can simplify the spin-observable local estimators as given below.
The diagonal spin-component estimators are given by
\begin{align}
    \sum_{\bsigma'} \braket{\bsigma | \hat{S}^x_i \hat{S}^x_j | \bsigma'} \frac{\psi(\bsigma')}{\psi(\bsigma)}
     & =
    \frac{1}{4}\frac{\psi(\bsigma^{(i,j)})}{\psi(\bsigma)},  \\
    \sum_{\bsigma'} \braket{\bsigma | \hat{S}^y_i \hat{S}^y_j | \bsigma'} \frac{\psi(\bsigma')}{\psi(\bsigma)}
     & =
    -\frac{1}{4}\sigma_i\sigma_j\frac{\psi(\bsigma^{(i,j)})}{\psi(\bsigma)}, \\
    \sum_{\bsigma'} \braket{\bsigma | \hat{S}^z_i \hat{S}^z_j | \bsigma'} \frac{\psi(\bsigma')}{\psi(\bsigma)}
     & =
    \frac{1}{4}\sigma_i\sigma_j,
\end{align}
where we used the fact that only the configuration with flipped spins at both sites $i$ and $j$ (i.e., $\bsigma' = \bsigma^{(i,j)}$) can contribute to the sums in the first two equations, while $\bsigma' = \bsigma$ is the only contributing configuration for the last equation.

The off-diagonal contributions read
\begin{align}
    \sum_{\bsigma'} \braket{\bsigma | \hat{S}^x_i \hat{S}^y_j | \bsigma'} \frac{\psi(\bsigma')}{\psi(\bsigma)}
     & =
    -\frac{\ri}{4}\sigma_j \frac{\psi(\bsigma^{(i,j)})}{\psi(\bsigma)}, \\
    \sum_{\bsigma'} \braket{\bsigma | \hat{S}^x_i \hat{S}^z_j | \bsigma'} \frac{\psi(\bsigma')}{\psi(\bsigma)}
     & =
    \frac{1}{4}\sigma_j \frac{\psi(\bsigma^{(i)})}{\psi(\bsigma)},      \\
    \sum_{\bsigma'} \braket{\bsigma | \hat{S}^y_i \hat{S}^z_j | \bsigma'} \frac{\psi(\bsigma')}{\psi(\bsigma)}
     & =
    -\frac{\ri}{4}\sigma_i\sigma_j \frac{\psi(\bsigma^{(i)})}{\psi(\bsigma)}.
\end{align}
The remaining contributions can be obtained by exchanging the spin indices $i$ and $j$ in the above equations.

Contributions involving single-spin operators are evaluated as
\begin{align}
    \sum_{\bsigma'} \braket{\bsigma | \hat{S}^x_i | \bsigma'} \frac{\psi(\bsigma')}{\psi(\bsigma)}
     & =
    \frac{1}{2}\frac{\psi(\bsigma^{(i)})}{\psi(\bsigma)},            \\
    \sum_{\bsigma'} \braket{\bsigma | \hat{S}^y_i | \bsigma'} \frac{\psi(\bsigma')}{\psi(\bsigma)}
     & =
    -\frac{\ri}{2}\sigma_i\frac{\psi(\bsigma^{(i)})}{\psi(\bsigma)}, \\
    \sum_{\bsigma'} \braket{\bsigma | \hat{S}^z_i | \bsigma'} \frac{\psi(\bsigma')}{\psi(\bsigma)}
     & =
    \frac{1}{2}\sigma_i.
\end{align}
These expressions provide local estimators for single-spin operators and can therefore be used to evaluate both local-energy contributions from Zeeman terms and observables such as local magnetization $m_i^\alpha = \braket{\hat{S}_i^\alpha}$.
For calculations in canonical local frames, Zeeman contributions are evaluated using transformed field coefficients $\tilde B_i^\alpha$. The physical magnetizations in the laboratory frame are subsequently reconstructed using \cref{eq:rotated_magnetization}.

\subsection{Energy Gradient}
Let us further assume the wavefunction is parametrized by a set of variational parameters $\btheta\in\mathbb R^{n_{\rm p}}$. In the search for a variational ground state, the parameters are optimized by minimizing the energy expectation value. The corresponding energy gradient is therefore a central ingredient of optimization and, as the energy itself, can be estimated from configurations sampled according to $p_{\btheta}(\bsigma)$.
The energy gradient with respect to the parameters $\btheta$ can be computed as
\begin{align}
    \partial_{\btheta}
    E(\btheta)
     & =
    \partial_{\btheta}
    \frac{\braket{\psi_{\btheta}|\hat H|\psi_{\btheta}}}
    {\braket{\psi_{\btheta}|\psi_{\btheta}}}
    \nonumber         \\
     & =
    2{\rm Re}
    \left(
    \frac{\braket{\partial_{\btheta}\psi_{\btheta}|\hat H|\psi_{\btheta}}}
    {\braket{\psi_{\btheta}|\psi_{\btheta}}}
    -
    \frac{\braket{\psi_{\btheta}|\hat H|\psi_{\btheta}}}
    {\braket{\psi_{\btheta}|\psi_{\btheta}}}
    \frac
    {\braket{\partial_{\btheta}\psi_{\btheta}|\psi_{\btheta}}}
    {\braket{\psi_{\btheta}|\psi_{\btheta}}}
    \right) \nonumber \\
     & \approx
    2{\rm Re}\left(
    \bbraket{E_{\rm loc}\bO^*}
    -
    \bbraket{E_{\rm loc}}
    \bbraket{\bO^*}
    \right) \nonumber \\
    & =
    2{\rm Re}\left(
    \bbraket{E^*_{\rm loc}\bO}
    -
    \bbraket{E^*_{\rm loc}}
    \bbraket{\bO}
    \right)
    \label{eq:loc_energy_O}
\end{align}
where we used the intermediary steps
\begin{align}
    \frac{\braket{\partial_{\btheta} \psi_{\btheta} | \hat H | \psi_{\btheta}}}{\braket{\psi_\btheta|\psi_\btheta}}
     & =
    \sum_{\bsigma, \bsigma'}
    \frac{\braket{\bsigma | \hat H | \bsigma'}}{\braket{\psi_\btheta|\psi_\btheta}}
    \braket{\partial_{\btheta}\psi_{\btheta} | \bsigma}
    \braket{\bsigma' | \psi_{\btheta}}
    \nonumber  \\
     & =
    \sum_{\bsigma, \bsigma'}
    \frac{\braket{\bsigma | \hat H | \bsigma'}}{\braket{\psi_\btheta|\psi_\btheta}}
    \psi_{\btheta}(\bsigma')
    (\partial_{\btheta} \psi_{\btheta}(\bsigma))^*
    \nonumber  \\
     & =
    \sum_{\bsigma, \bsigma'}
    \frac{\braket{\bsigma | \hat H | \bsigma'}}{\braket{\psi_\btheta|\psi_\btheta}}
    \psi_\btheta(\bsigma')
    \psi^*_\btheta(\bsigma)
    (\partial_{\btheta} \ln\psi_{\btheta}(\bsigma))^*
    \nonumber  \\
     & =
    \sum_{\bsigma, \bsigma'}
    \frac{|\psi_{\btheta}(\bsigma)|^2}{\braket{\psi_\btheta|\psi_\btheta}}
    \braket{\bsigma | \hat H | \bsigma'}
    \frac{\psi_\btheta(\bsigma')}{\psi_\btheta(\bsigma)}
    \boldsymbol O^*(\bsigma)
    \nonumber  \\
     & =
    \sum_{\bsigma}
    p_\btheta(\bsigma)
    E_{\rm loc}(\bsigma) \boldsymbol O^*(\bsigma)
    \nonumber  \\
     & \approx
    \bbraket{E_{\rm loc}\boldsymbol O^*}
    ,
    \label{eq:energy_gradient}
\end{align}
used the sample average defined in \cref{eq:sample_average} and defined the logarithmic wavefunction derivative $\boldsymbol O(\bsigma) \in \mathbb{C}^{n_{\rm p}}$. Component-wise, the logarithmic derivative reads $O_i(\bsigma) = \partial_{\theta_i}\ln\psi_\btheta(\bsigma) = \partial_{\theta_i}\psi_{\btheta}(\bsigma)/\psi_{\btheta}(\bsigma)$.

\section{Sampling}
\label{sec:sampling}

\subsection{Markov-chain Monte Carlo}

Expectation values are estimated by sampling configurations $\bsigma$ from the variational probability distribution $p_{\btheta}(\bsigma)=|\psi_{\btheta}(\bsigma)|^2 /\braket{\psi_{\btheta}|\psi_{\btheta}}.$
The Hilbert space grows exponentially with system size, making direct summation over all configurations intractable. We therefore employ Markov-chain Monte Carlo (MCMC) to generate samples distributed according to $p_{\btheta}(\bsigma)$.
We construct a Markov chain with transition probability $t(\bsigma\to\bsigma')$ that satisfies a detailed balance $p_{\btheta}(\bsigma)\,t(\bsigma\to\bsigma')=p_{\btheta}(\bsigma')\,t(\bsigma'\to\bsigma)$ such that the stationary distribution of the chain is $p_{\btheta}(\bsigma)$.

We employ the Metropolis algorithm with symmetric proposals. Given a current configuration $\bsigma$, a trial configuration $\bsigma'$ is proposed and accepted with probability
\begin{equation}
    A(\bsigma\to\bsigma')
    =
    \min\left[
        1,
        \frac{p_{\btheta}(\bsigma')}
        {p_{\btheta}(\bsigma)}
        \right]
    =
    \min\left[
        1,
        \frac{|\psi_{\btheta}(\bsigma')|^2}
        {|\psi_{\btheta}(\bsigma)|^2}
        \right].
    \label{eq:metropolis_accept}
\end{equation}
For numerical stability, we evaluate the probability ratio using the logarithm of the wavefunction $\ln\psi_{\btheta}(\bsigma)$,
\begin{equation}
    \frac{|\psi_{\btheta}(\bsigma')|^2}{|\psi_{\btheta}(\bsigma)|^2}
    =
    \exp\left[
        2{\rm Re}\left(\ln\psi_{\btheta}(\bsigma')-\ln\psi_{\btheta}(\bsigma)\right)
        \right].
\end{equation}

\subsection{Proposal move and sampling protocol}

We use local single-spin-flip proposals. For a configuration $\bsigma\in\{\pm1\}^{N}$, a site $i\in\{1,\dots,N\}$ is selected uniformly at random, and the trial configuration $\bsigma'$ is obtained by flipping $\sigma_i\to-\sigma_i$.
Thus, the proposal probability $q(\bsigma'|\bsigma)$, defined as the probability of proposing $\bsigma'$ given the current configuration $\bsigma$, is $1/N$ for configurations $\bsigma'$ that differ from $\bsigma$ by a single spin flip and zero otherwise.
Since the spin flip is reversible, the proposal distribution is symmetric $q(\bsigma'|\bsigma)=q(\bsigma|\bsigma')$, and therefore the acceptance probability is given by \cref{eq:metropolis_accept}.

At the beginning of optimization, the $n_{\rm batch}$ chains are initialized from random spin configurations, with each spin chosen independently and uniformly from $\{-1,1\}$. Sampling proceeds immediately, without a burn-in period (i.e., a waiting period to allow the chains to equilibrate towards the stationary distribution of the target distribution).
The Markov chains are then evolved independently and persistently across optimization steps, with the target distribution $p_{\btheta}(\bsigma)$ updated at each step.
Between successive optimization steps, each chain evolves for $n_{\rm res}$ Metropolis updates. Subsequently, the final configuration is retained from each chain, giving $n_s = n_{\rm batch}$ samples per iteration. The resulting configurations are then used to evaluate local estimators as given in \cref{sec:matrix_elements}.

For the simulations reported in the manuscript, we choose $n_{\rm res}=N$, which corresponds to one attempted spin update per site on average between successive optimization steps. All chains and Metropolis updates are evaluated in parallel over the batch dimension and the batch size is chosen as $n_{\rm batch}=2^{19}$.

\section{Stochastic Reconfiguration (SR)}
\label{sec:SR}

Let us assume that the states are parameterized by a set of real parameters $\btheta$. Stochastic reconfiguration can be interpreted as a natural-gradient update on the variational manifold of quantum states $\ket{\psi_\btheta}$.
The projective Hilbert space is a Kähler manifold endowed with the Fubini-Study metric and a compatible symplectic form. The variational family is a real submanifold that inherits a Riemannian metric $g$ and a two-form $w$ from this structure, given by $g=\mathrm{Re}Q$ and $w=-2\mathrm{Im}Q$, where $Q$ is the quantum geometric tensor pulled back to the variational submanifold. The submanifold is generally not itself Kähler, since it need not inherit the compatible complex structure.
We induce the metric $g$ from the Fubini-Study quantum distance
\begin{align}
    d_{\rm FS}^2
     & =
    -\ln\left(
    \frac{
        |\braket{\psi_\btheta | \psi_{\btheta'}}|^2
    }{
        \braket{\psi_\btheta | \psi_\btheta}
        \,
        \braket{\psi_{\btheta'} | \psi_{\btheta'}}
    }
    \right)
    .
    \label{eq:induced_metric}
\end{align}
Assuming $\btheta' = \btheta + \delta\btheta$, the argument can be Taylor expanded up to second order in $\delta\btheta$ as
\begin{align}
    \frac{
        |\braket{\psi_\btheta | \psi_{\btheta'}}|^2
    }{
        \braket{\psi_\btheta | \psi_\btheta}
        \,
        \braket{\psi_{\btheta'} | \psi_{\btheta'}}
    }
     & \approx
    \left(
    1 + a_i\delta \theta_i
    +
    \left(b_{ij} -
        {\rm Re}Q_{ij}\right)
    \delta \theta_i \delta \theta_j
    \right)
    \nonumber  \\
     & \times
    \left(
    1 -
    a_i\delta \theta_i
    +
    \left(a_i a_j - b_{ij}\right)\delta \theta_i\delta \theta_j
    \right)
    \nonumber  \\
     & \approx
    1 - {\rm Re}Q_{ij}\delta \theta_i\delta \theta_j
    ,
\end{align}
where the summation convention is assumed.
The above drifting-norm terms $a_i = \partial_i\ln\braket{\psi_\btheta | \psi_\btheta}$ and $b_{ij} = \frac{1}{2\braket{\psi_\btheta | \psi_\btheta}}\partial_i\partial_j \braket{\psi_\btheta | \psi_\btheta}$ out, resulting in the infinitesimal Fubini-Study distance
\begin{align}
    d_{\rm FS}^2
     & \approx -\ln{(1- {\rm Re}Q_{ij}\delta \theta_i\delta \theta_j)}
    \nonumber                                                \\
     & \approx {\rm Re}Q_{ij}\delta \theta_i\delta \theta_j,
\end{align}
where $Q_{ij}$ represent the elements of the quantum geometric tensor
\begin{align}
    Q_{ij} =
    \frac1{\braket{\psi_\btheta | \psi_\btheta}}
    \Bra{
        \frac{\partial\psi_\btheta}{\partial \theta_i} }
    \left(1 - \hat P
    \right)
    \Ket{\frac{\partial\psi_\btheta}{\partial \theta_j}},
    \label{eq:quantum_geom_tensor}
\end{align}
and $\hat P$ is the state projector;
\begin{equation}
    \hat P
    =
    \frac{\ket{\psi_\btheta}\bra{\psi_\btheta}}
    {\braket{\psi_\btheta | \psi_\btheta}}.
    \label{eq:state_projector}
\end{equation}
It follows that for infinitesimally close states, the Fubini-Study distance induces a Riemannian metric on the parameter space given as $g_{ij} = {\rm Re} Q_{ij}$.

We further define the (complex) Berry connection
\begin{equation}
    \bBC = \ri\frac{\braket{\psi_\btheta | \partial_\btheta\psi_\btheta}}{\braket{\psi_\btheta | \psi_\btheta}}.
\end{equation}
By straightforward evaluation, we find $Q_{ij} = Q_{ji} - \ri \left(\partial_i \BC_j - \partial_j \BC_i\right)$, from which we deduce that the Berry curvature $w_{ij}=\left(\partial_i \BC_j - \partial_j \BC_i\right)$ is the anti-symmetric two-form.
The Berry connection motivates us to define the gauge-covariant derivative $\bnabla \coloneq \partial_{\btheta} + \ri\bBC$, whose commutator is the curvature $w_{ij} = -\ri [\nabla_i, \nabla_j]$.
Under the transformation $\ket{\psi_\btheta}\rightarrow \re^{\ri\varphi(\btheta)}\ket{\psi_\btheta}$, we find
\begin{align}
    \BC_i
     & =
    \ri
    \braket{\psi_\btheta | \partial_i | \psi_\btheta} /
    \braket{ \psi_\btheta | \psi_\btheta}
    \rightarrow
    \nonumber \\
    & \rightarrow
    \ri
    \braket{\psi_\btheta | \re^{-\ri\varphi(\btheta)} \partial_i \re^{\ri\varphi(\btheta)} | \psi_\btheta}
    /\braket{ \psi_\btheta | \psi_\btheta}
    \nonumber      \\
    &=
    \BC_i
    -
    \partial_i
    \varphi(\btheta),
\end{align}
and further
\begin{align}
    \bnabla\ket{\psi_\btheta}
     &
    \rightarrow
    (\partial_\btheta + \ri\bBC - \ri\partial_\btheta\varphi(\btheta))\re^{\ri\varphi(\btheta)}\ket{\psi_\btheta}
    \nonumber \\
     & =
    \re^{\ri\varphi(\btheta)}\bnabla\ket{\psi_\btheta}
    \label{eq:covariant_derivative_gauge_transformation}
\end{align}
Therefore, the covariant derivative transforms covariantly under gauge transformations, and we find
\begin{align}
    \frac{
    \braket{\psi_\btheta | \bnabla | \psi_\btheta}}
    {\braket{\psi_\btheta | \psi_\btheta}}
     & =
    \frac{
    \braket{\psi_\btheta | \partial_\btheta | \psi_\btheta}}
    {\braket{\psi_\btheta | \psi_\btheta}}
    +
    \ri\bBC
    =
    0
    \label{eq:tangent_states_orthogonality}
    .
\end{align}
The quantum geometric tensor can be compactly expressed in terms of the covariant derivative as
\begin{align}
    Q_{ij} =
    \frac{
    \braket{\nabla_i\psi_\btheta | \nabla_j\psi_\btheta}}
    {\braket{\psi_\btheta | \psi_\btheta}}
    ,
\end{align}
which highlights its gauge independence due to \cref{eq:covariant_derivative_gauge_transformation}.
Finally, we can equip every point of the manifold with a tangent space $T_\btheta M = {\rm span}\{\ket{\bnabla\psi_\btheta}\}$, endowed with the inner product $\langle u|v\rangle_{T_\btheta M} := \langle u|v\rangle / \braket{\psi_\btheta|\psi_\btheta}$.

In order to evolve an initial state with finite ground state overlap closer towards the ground state, we apply imaginary time-evolution, i.e.
\begin{align}
    \ket{\phi_\btheta}
    =
    {\rm e}^{-2\eta\hat H}\ket{\psi_{\btheta}}
    .
\end{align}
To find the optimal tangential vector, we extremize the (in this case gauge-invariant) vector distance
\begin{align}
    d^2_{\rm vec}
    =
    |\ket{t_\btheta} - \ket{\phi_\btheta}|^2/\braket{\psi_\btheta | \psi_\btheta}
    \label{eq:vector_distance}
\end{align}
with respect to the coordinates $c_k$ of the tangent state $\ket{t_\btheta} = c_i\nabla_i\ket{\psi_\btheta}$. We restrict the coefficients to be real ${\bm c} \in \mathbb{R}$, as appropriate for a variational family with real parameters $\btheta$.
With the tangent-space inner product, we have $\braket{t_\btheta | t_\btheta}_{T_\btheta M} = c_i c_j g_{ij}$.
Requiring $\partial_{c_k}d^2_{\rm vec}=0$ leads us to
\begin{align}
    g_{kj} c_j = \frac{{\rm Re}\braket{\psi_\btheta | \re^{-2\eta\hat H} \nabla_k | \psi_\btheta}}{\braket{\psi_\btheta|\psi_\btheta}}
    .
\end{align}
For sufficiently small $\eta>0$ we can approximate $\re^{-2\eta\hat H} \approx 1 - 2\eta\hat H$.
Since $\nabla_i\ket{\psi_\btheta}$ is orthogonal to $\ket{\psi_\btheta}$ in the ambient Hilbert space (see \cref{eq:tangent_states_orthogonality}), the leading-order term vanishes and we find
\begin{align}
    g_{kj}c_j
     & \approx
    -2\eta \frac{{\rm Re}\braket{\psi_\btheta | \hat H \nabla_k | \psi_\btheta}}{\braket{\psi_\btheta|\psi_\btheta}}
    \nonumber  \\
     & =
    -\eta
    2{\rm Re}
    \left(
    \frac{\braket{\psi_\btheta | \hat H | \partial_k \psi_\btheta}}{\braket{\psi_\btheta|\psi_\btheta}}
    -
    \frac{\braket{\psi_\btheta | \hat H | \psi_\btheta}}{\braket{\psi_\btheta|\psi_\btheta}}
    \frac{\braket{\psi_\btheta | \partial_k \psi_\btheta}}{\braket{\psi_\btheta|\psi_\btheta}}
    \right)
    \nonumber  \\
     & =
    -\eta \partial_k E
    .
\end{align}
If we now require that the entire right-hand side $\eta g_{jk}^{-1}\partial_k E$ be small enough to approximate $\ket{\psi_{\btheta + \delta\btheta}} \approx \ket{\psi_\btheta} + \delta\btheta \cdot \bnabla\ket{\psi_\btheta}$, we can identify $c_j = \delta \theta_j$ and arrive at
\begin{align}
    g_{ij}\delta \theta_j = -\eta \partial_i E
    ,
    \label{eq:stochastic_reconfiguration}
\end{align}
which in the Monte Carlo community is the defining equation of stochastic reconfiguration (SR).
We can now see that \cref{eq:stochastic_reconfiguration} takes the form of a Riemannian gradient descent, taking into account the metric distance in the manifold of variational states.

\section{Regularized Riemannian Gradient Descent}
\label{sec:Apdx_RRGD}

\subsection{Derivation of constrained gradient descent}

For real parameters $\btheta\in\mathbb R^{n_{\rm p}}$, consider the minimization of the locally linearized energy subject to a fixed Euclidean step,
\begin{equation}
    \mathcal L_{\rm GD}(\delta\btheta)
    =
    \partial_{\btheta}E\cdot\delta\btheta
    +
    \frac{\lambda}{2}
    \left(
    |\delta\btheta|^2-\eta'^2
    \right).
\end{equation}
Stationarity and the norm constraint give
\begin{equation}
    \delta\btheta
    =
    -\eta'
    \frac{\partial_{\btheta}E}
    {|\partial_{\btheta}E|}.
\end{equation}
Hence, gradient descent is locally the direction that minimizes the energy at fixed Euclidean distance.

For variational quantum states, replacing the Euclidean norm by the quantum line element gives
\begin{equation}
    \mathcal L_{\rm RGD}(\delta\btheta)
    =
    \partial_{\btheta}E\cdot\delta\btheta
    +
    \frac{\lambda}{2}
    \left(
    \delta\btheta^{T}g\delta\btheta-\eta'^2
    \right).
\end{equation}
Assuming that $g$ is invertible in the relevant tangent subspace, this yields
\begin{equation}
    g\delta\btheta
    =
    -\frac{1}{\lambda}
    \partial_{\btheta}E,
    \qquad
    \lambda
    =
    \frac{
        \sqrt{
            (\partial_{\btheta}E) \cdot
            (g^{-1}
            \partial_{\btheta}E)
        }
    }{\eta'}.
    \label{eq:rgd_lambda}
\end{equation}
Instead of a fixed $\eta'$, it is more convenient to absorb the gradient norm into the learning rate, i.e. $\eta=\eta'/\sqrt{\partial_{\bm\theta}E(\bm\theta)\cdot (g^{-1}\partial_{\bm\theta}E(\bm\theta))}$, such that the update step simplifies to the SR equation \cref{eq:stochastic_reconfiguration}.
This has the benefits that (i) a constant $\eta$ dynamically adapts the update step's quantum distance (equal to $\eta'$) and (ii) we do not need to invert the large dense matrix representation of the quantum metric $g$ but may rely on matrix-free conjugate gradient methods to find a non-trivial solution to \cref{eq:stochastic_reconfiguration}.

To additionally bound the endpoint in parameter space, consider
\begin{equation}
    |\btheta+\delta\btheta|^2\leq\mu'^2.
\end{equation}
The associated Karush-Kuhn-Tucker Lagrangian is
\begin{align}
    \mathcal L_{\rm RRGD}(\delta\bm\theta)
    =
    \mathcal L_{\rm RGD}(\delta\bm\theta)
    +
    \frac\nu2
    \left(|\bm\theta + \delta\bm\theta|^2 - \mu'^2\right)
    ,
\end{align}
with $\nu\geq0$ and complementary slackness condition $\nu\left(|\btheta+\delta\btheta|^2-\mu'^2\right)=0$. 
The inequality constraint is enforced only when the updated parameters reach the boundary of the allowed region, in which case $\nu$ may be nonzero. If the constraint is inactive, i.e., the updated parameters lie strictly within the allowed region, complementary slackness requires ($\nu=0$).
An exactly constrained update would require solving a nonlinear system of equations for the multipliers at each iteration.
Instead, we relax the hard constraint and treat $\eta=1/\lambda$ and $\mu=\nu/\lambda$ as hyperparameters, obtaining
\begin{equation}
    (g+\mu\mathbbm 1)\delta\btheta
    =
    -\eta\,\partial_{\btheta}E-\mu\btheta.
    \label{eq:RRGD_sup}
\end{equation}
This defines the RRGD equation. The resulting RRGD update softly penalizes the endpoint norm rather than enforcing a prescribed value of $\mu'$.

\subsection{Spectral properties and regularization bias}

We analyze the action of RRGD in the eigenbasis of the quantum metric. Let
\begin{equation}
    g\bm v_a=\gamma_a\bm v_a,
    \qquad
    \gamma_a\geq0,
\end{equation}
and define $\theta_a=\bm v_a^{T}\btheta$ and $\partial_aE=\bm v_a^{T}\partial_{\btheta}E$. Projecting \cref{eq:RRGD_sup} onto $\bm v_a$ gives
\begin{equation}
    \delta\theta_a
    =
    -\frac{\eta\,\partial_aE+\mu\theta_a}
    {\gamma_a+\mu}
    =
    -\frac{\eta}{\gamma_a+\mu}\partial_aE
    -
    \frac{\mu}{\gamma_a+\mu}\theta_a .
    \label{eq:RRGD_eigenmode}
\end{equation}
Although the Euclidean penalty is isotropic before preconditioning, its effect on the update is therefore filtered by the quantum metric. For $\gamma_a\gg\mu$, the energy-driven term approaches the SR update,
\begin{equation}
    \delta\theta_a
    =
    -\frac{\eta}{\gamma_a}\partial_aE
    +\mathcal O\left(\frac{\mu}{\gamma_a}\right),
\end{equation}
and the restoring contribution is suppressed by $\mu/\gamma_a$. In contrast, the directions with $\gamma_a\lesssim\mu$ are both regularized and driven toward the reference parameters.

This selectivity is physically motivated by the relation between the metric and the energy gradient. For an unnormalized state,
\begin{equation}
    \partial_aE
    =
    \frac{2}{\braket{\psi | \psi}}\,{\rm Re}
    \Braket{
        \partial_a\psi
        |
        (\hat H-E)
        |
        \psi
    } .
\end{equation}
Since $\braket{\psi | (\hat H-E)| \psi} = 0$, it equivalently holds that
\begin{equation}
    \partial_aE
    =
    \frac{2}{\braket{\psi | \psi}}\,{\rm Re}
    \Braket{
        \partial_a\psi
        |
        (1 - \hat P)
        (\hat H-E)
        |
        \psi
    } .
\end{equation}
where $\hat P$ is the projector from \cref{eq:state_projector}.
The Cauchy--Schwarz inequality gives
\begin{align}
    |\partial_aE|
     & \leq
    2\sqrt{(\bm v_a)_i (\bm v_a)_j Q_{ij}}
    \sqrt{ \sigma^2 } \nonumber \\
     & =
    2\sqrt{{\bm v}_a^T g \bm v_a }
    \sqrt{ \sigma^2 } \nonumber \\
     & =
    2\sqrt{\gamma_a}
    \sqrt{ \sigma^2 },
    \label{eq:gradient_metric_bound}
\end{align}
where we identified the quantum geometric tensor $Q$ from \cref{eq:quantum_geom_tensor} and the variance of energy $\sigma^2 = \Braket{\psi|(\hat H-E)^2|\psi} / \braket{\psi | \psi}$.
Thus, an exact null direction of the metric cannot carry an exact energy gradient. For small but finite $\gamma_a$, the directional energy derivative is correspondingly bounded, while the SR update divides this derivative by $\gamma_a$. Inverting the metric can therefore produce large parameter displacements and amplify sampling errors along directions to which the state is only weakly sensitive. RRGD limits this amplification by replacing $\gamma_a^{-1}$ with $(\gamma_a+\mu)^{-1}$ and simultaneously anchors the corresponding parameter component.

The parameter anchoring can be viewed as a soft gauge fixing. Redundant parameterizations may represent the same normalized state at different points in parameter space; the quantum metric alone does not select among such equivalent representations. The restoring term favors the small-norm representative centered on the chosen reference point. More generally, anchoring around $\btheta_{\rm ref}$ gives
\begin{equation}
    (g+\mu\mathbbm 1)\delta\btheta
    =
    -\eta\,\partial_{\btheta}E
    -\mu(\btheta-\btheta_{\rm ref}).
    \label{eq:RRGD_reference}
\end{equation}
In the canonical local frames used here, $\btheta_{\rm ref}=\bm0$ represents the physically motivated reference texture, reducing \cref{eq:RRGD_reference} to \cref{eq:RRGD_sup}.

A finite anchoring strength can also bias the optimized state. Defining $\nu=\mu/\eta$, the update can be written as
\begin{equation}
    (g+\eta\nu\mathbbm 1)\delta\btheta
    =
    -\eta\,
    \partial_{\btheta}
    \left[
        E(\btheta)+\frac{\nu}{2}|\btheta|^2
        \right].
    \label{eq:RRGD_regularized_objective}
\end{equation}
Consequently, for fixed $\eta$ and $\mu$, an RRGD fixed point generally satisfies
\begin{equation}
    \partial_{\btheta}E
    =
    -\frac{\mu}{\eta}\btheta,
\end{equation}
rather than $\partial_{\btheta}E=\bm0$. The ratio $\mu/\eta$ therefore controls the strength of the parameter-norm penalty in the optimization objective. For a fixed learning rate $\eta$, ordinary SR is formally recovered as $\mu/\eta\rightarrow0$, whereas keeping $\mu=\eta\nu$ finite corresponds to optimizing with a fixed $L_2$ penalty of strength $\nu$.

Note that regularization modifies the optimization objective, but not the Hamiltonian expectation value $E(\btheta)$ evaluated at a given state. Its effect on the physical energy is indirect: the penalty favors small-norm parameters and, therefore, changes the state selected by optimization. 
Consequently, the penalty value $\nu|\btheta|^2/2$, with $\nu=\mu/\eta$, is not the difference between the regularized and unregularized minimum energies. Subtracting it from the measured energy, therefore, does not, in general, recover the unregularized variational minimum. 
To reduce this bias, the anchoring strength $\nu$ can gradually be decreased, allowing the state to reoptimize. This procedure approaches unregularized optimization but does not eliminate errors by itself due to incomplete convergence, local minima, or limited variational expressivity.

\subsection{Convergence and numerical stability}
\begin{figure*}
    \includegraphics{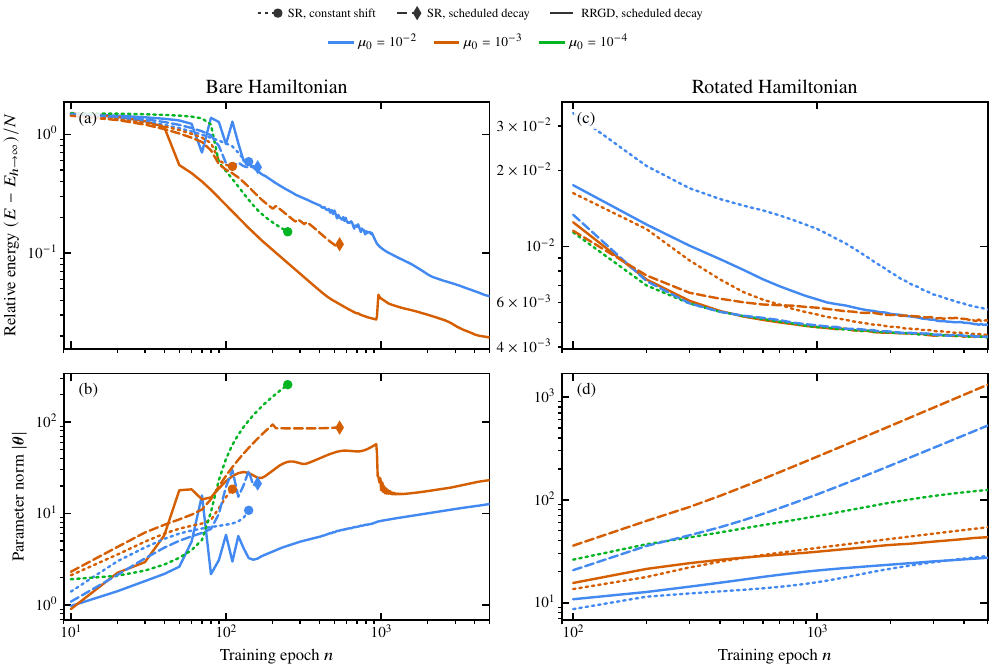}
    \caption{
        Comparison of SR and RRGD for the bare Hamiltonian (left column) and the rotated Hamiltonian (right column).
        (a,c)~Energy difference per spin relative to the reference energy; the infinite-head extrapolation $E_{h\to\infty}$.
        (b,d)~Euclidean norm of the variational parameters.
        Colors indicate the initial regularization strength $\mu_0$; dotted, dashed, and solid lines denote SR with constant shift, SR with scheduled decay, and RRGD with scheduled decay, respectively. Circles and diamonds mark the final recorded iterations of SR runs
        that terminate because of NaN energy estimates.
        A rollback procedure restores an earlier checkpoint and increases the diagonal regularization according to \cref{eq:restoring_diag_reg}, whenever the update fails the prescribed stability criteria. 
        No rollback is triggered in the rotated-frame runs.
    }
    \label{fig:method_comparisson}
\end{figure*}

To compare the SR and RRGD methods, we employ the antiferromagnetic triangular monolayer introduced in the main text. We train a two-headed network with the Hamiltonian \eqref{eq:hamiltonian} with $J=1$, $B=-1.76$, N\'eel-type DMI with strength $D_{\rm N}=0.8$ and periodic boundaries. The training methods are compared for both the bare Hamiltonian $H$ and the rotated Hamiltonian $H'$, considering different initial regularization strengths $\mu_0$. The results are shown in \cref{fig:method_comparisson}.
As discussed above, it follows from \cref{eq:RRGD_regularized_objective} that a finite regularization strength $\mu$ introduces a bias in the energy landscape. We therefore employ a scheduled decay of $\mu$ using the sequence 
\begin{equation}
    \mu_{n+1} 
    = 
    \max\left(\mu_n \frac{n+1}{n+2}, 10^{-7}\right),
    \label{eq:sched_decay}
\end{equation}
where $n$ denotes the training step (epoch). This decay is applied after each accepted update.

To prevent premature under-regularization and mitigate numerical instabilities, we additionally implement a rollback procedure based on the size and finiteness of the parameter update. 
Let $\delta\btheta_n$ denote the change in the parameter vector during an optimization epoch $n$. During the first $n_{\rm w}=20$ ``warm-up'' epochs, a step is accepted provided that $|\delta\btheta_n|$ is finite. 
Thereafter, we also require the relative update norm to satisfy
\begin{equation}
    \frac{|\delta\btheta_n|}{|\btheta_{n-1}|+\epsilon_0} < R,
\end{equation}
where $R=5$ and $\epsilon_0=10^{-12}$. Once at least ten accepted update norms are available, we impose the additional condition
\begin{equation}
|\delta\btheta_n|
<
10 \overline{\Delta}_n,
\qquad
\overline{\Delta}_n
=
\frac{1}{10}\sum_{j=1}^{10}|\delta\btheta_{n_j}|,
\end{equation}
where the average is taken over the ten most recent accepted updates, which need not correspond to consecutive epochs.

If a step fails any of these criteria, the parameter vector is restored to the oldest checkpoint in a buffer containing up to 20 accepted parameter vectors, and the regularization strength is increased according to
\begin{equation}
    \mu_n\rightarrow\min(100\mu_n,\mu_0).
    \label{eq:restoring_diag_reg}
\end{equation}
The update history is retained, and the regularization schedule resumes after the next accepted step according to \cref{eq:sched_decay}. Independently, training terminates if the energy estimate becomes NaN.

We observe that for the rotated Hamiltonian, SR and RRGD reach comparable energies over the given iteration budget [\cref{fig:method_comparisson}(c)]. 
Their parameter norms, however, evolve differently: SR with scheduled decay exhibits a substantially stronger growth than RRGD, without a corresponding improvement in energy [\cref{fig:method_comparisson}(d)]. A constant diagonal shift suppresses this growth, although the larger shift also slows energy convergence. 
All runs in the rotated frame remain numerically stable, and none trigger the rollback procedure described above.

For the bare unrotated Hamiltonian, all SR runs displayed eventually terminate because of NaN energy estimates, as indicated by their endpoint markers [\cref{fig:method_comparisson}(a,b)]. Their parameter norms exhibit pronounced growth or irregular variations before termination.
Repeated rollback attempts do not restore sustained convergence in these runs, and they ultimately fail. Thus, among the configurations here, only RRGD combined with rollback sustains stability throughout the displayed interval.

These observations are consistent with the stabilizing role of the explicit restoring term $-\mu\btheta$. RRGD maintains comparatively controlled parameter norms while reaching energies comparable to SR in the rotated frame and sustaining energy reduction in the bare frame over the displayed training interval.
The results therefore demonstrate improved numerical robustness for the tested configurations, although they do not establish convergence to a stationary point or show that RRGD is unconditionally stable. The effectiveness of the method remains dependent on the regularization strength and the rollback procedure.

\section{Linear solvers}

Analogously to \cref{eq:loc_energy_O}, using the sample average defined in \cref{eq:sample_average}, we can express the quantum metric in terms of the logarithmic wavefunction derivatives as
\begin{align}
    g_{ij}
     & =
    {\rm Re}\left(
    \frac{\braket{\partial_i \psi_\btheta | \partial_j \psi_\btheta}}{\braket{\psi_\btheta|\psi_\btheta}}
    -
    \frac{\braket{\partial_i \psi_\btheta | \psi_\btheta}}{\braket{\psi_\btheta|\psi_\btheta}}
    \frac{\braket{\psi_\btheta | \partial_j \psi_\btheta}}{\braket{\psi_\btheta|\psi_\btheta}}
    \right)
    \nonumber \\
     & =
    {\rm Re}\left(\bbraket{O^*_i O_j} - \bbraket{O^*_i} \bbraket{O_j}
    \right),
\end{align}
where we used that
\begin{align}
    \frac{\braket{\partial_i \psi_\btheta | \partial_j \psi_\btheta}}{\braket{\psi_\btheta|\psi_\btheta}}
     & =
    \sum_{\bsigma} \frac{1}{\braket{\psi_\btheta | \psi_\btheta}}
    \braket{ \partial_i \psi_\btheta | \bsigma}
    \braket{\bsigma | \partial_j \psi_\btheta}
    \nonumber  \\
     & =
    \sum_{\bsigma} \frac{1}{\braket{\psi_\btheta | \psi_\btheta}}
    \left(\partial_i \psi_\btheta (\bsigma)\right)^*
    \partial_j \psi_\btheta (\bsigma)
    \nonumber  \\
     & =
    \sum_{\bsigma} \frac{\vert\psi_\btheta(\bsigma)\vert^2}{\braket{\psi_\btheta | \psi_\btheta}}
    \left(\partial_i \ln\psi_\btheta (\bsigma) \right)^*
    \partial_j \ln\psi_\btheta (\bsigma)
    \nonumber  \\
     & =
    \sum_{\bsigma} p_\btheta(\bsigma) O^*_i(\bsigma) O_j(\bsigma)
    \nonumber  \\
     & \approx
    \bbraket{O^*_i O_j}.
\end{align}

To avoid forming full matrices, we instead compute Jacobian-vector (JVP) and vector-Jacobian (VJP) products on the fly, which avoids the explicit construction of the $n_s\times n_{\rm p}$ Jacobian or the $n_{\rm p}\times n_{\rm p}$ metric and therefore keeps the memory at that of a forward and backward pass over the batch \cite{Novak2022}.
The JVP and VJP exploit the forward- and reverse-mode autodiff primitives of the JAX library \cite{jax2018github}, respectively. They are defined as the contraction of a (co)vector $\bm v$ with the Jacobian $\partial f_i(\bm x)/\partial \theta_j$ of the multivariate function $\bm f$ with components $f_i(\bm x)$, i.e.:
\begin{align}
    {\rm JVP}_{\bm f}(\bm v) & = \frac{\partial f_i(\bm x)}{\partial \theta_j} v_j. \\
    {\rm VJP}_{\bm f}(\bm v) & = v_i\frac{\partial f_i(\bm x)}{\partial \theta_j}
\end{align}
Recast in terms of these operations, the contraction of the quantum metric with a vector $\delta\btheta$ can be expressed as
\begin{align}
    g\delta\btheta
     & =
    {\rm Re}
    \left(
    \frac{1}{n_s}
    {\rm VJP}_{(\ln \bm\psi)^*}({\rm JVP}_{\ln\bm\psi}(\delta\btheta))
    \right)-
    \nonumber \\
     & -
    {\rm Re}
    \left(
    {\rm VJP}_{\bbraket{\ln \psi}^*}({\rm JVP}_{\bbraket{\ln\psi}}(\delta\btheta))
    \right)
    ,
\end{align}
and the energy gradient \eqref{eq:energy_gradient} reads
\begin{align}
    \partial_{\bm\theta}E(\bm\theta)
    &=
    2{\rm Re}\left(
        \frac1{n_s}{\rm VJP}_{\ln \bm\psi}(\bm E^*_{\rm loc})  \right) -
    \nonumber\\
    &-
    2{\rm Re}\left(
        {\rm VJP}_{\bbraket{\ln\psi}}(\bbraket{E^*_{\rm loc}})
        \right),
\end{align}
where $\ln\bm\psi$ and $\bm E_{\rm loc}$ are vectors with components $\ln\psi_i = \ln\psi(\bm\sigma_i)$ and $E_{{\rm loc},i} = E_{\rm loc}(\bm\sigma_i)$, respectively.

This matrix-free implementation requires only action on trial vectors, which is suitable for iterative linear solvers. 
In this work, we use the conjugate gradient method implemented in the JAX library \cite{jax2018github} as \texttt{jax.scipy.sparse.linalg.cg}.